\PassOptionsToPackage{numbers}{natbib}
\documentclass[]{Liebert_Author}

\usepackage{authblk}
\usepackage{graphicx}
\usepackage[version=4]{mhchem}
\usepackage[english]{babel}
\usepackage[utf8]{inputenc}
\usepackage[margin=1in]{geometry}
\usepackage{amsmath}
\usepackage{csquotes}

\def\perK{~K$^{-1}$}
\def\Wm2K{~W~m$^{-2}$\perK}

\title{Strong Evidence That Abiogenesis Is \\[-1.5ex]a Rapid Process on Earth Analogs} 

\author[1,*]{David Kipping}
\affil[1]{Dept. of Astronomy, Columbia University, 550 W 120th St., New York, NY 10027}
\date{}

\begin{document} 

\maketitle 

\begin{abstract}
The early start to life naively suggests that abiogenesis is a rapid process on Earth-like planets. However, if evolution typically takes ${\sim}4$\,Gyr to produce intelligent life-forms like us, then the limited lifespan of Earth's biosphere (${\sim}$5-6\,Gyr) necessitates an early (and possibly highly atypical) start to our emergence - an example of the weak anthropic principle. Our previously proposed objective Bayesian analysis of Earth’s chronology culminated in a formula for the minimum odds ratio between the fast and slow abiogenesis scenarios (relative to Earth’s lifespan). Timing from microfossils (3.7\,Gya) yields 3:1 odds in favor of rapid abiogenesis, whereas evidence from carbon isotopes (4.1\,Gya) gives 9:1, both below the canonical threshold of ``strong evidence'' (10:1). However, the recent result of a 4.2\,Gya LUCA pushes the odds over the threshold for the first time (nominally 13:1). In fact, the odds ratio is ${>}$10:1 for all possible values of the biosphere's ultimate lifespan and speculative hypotheses of ancient civilizations. For the first time, we have formally strong evidence that favors the hypothesis that life rapidly emerges in Earth-like conditions (although such environments may themselves be rare).

\end{abstract}

\section{The Inference Hurdle Posed by the Weak Anthropic Principle}

The earliest fossil evidence for life on Earth comes from 3.7\,Gyr old metamorphosed sedimentary rocks in southwest Greenland, which contain 1-4\,cm high stromatolites \cite{nutman:2016}. Earth plausibly became suitable for the emergence of life once the oceans formed $(4.404\pm0.008)$\,Gya \cite{wilde:2001} and thus life clearly began quickly ($\leq700$\,Myr). It is tempting to assume that this timescale would be typical on similar Earth-like exoplanets and therefore that ``Life is not a fussy, reluctant and unlikely thing'' to quote a commentary article associated with the stromatolite discovery \cite{allwood:2016}.

Consider, however, the possibility that the timescale for evolution to run its course and produce self-aware organisms capable of statistics, geology, palaeontology, and so on\footnote{What one might define as ``intelligence'' for the purposes of this article, although it is acknowledged that this is an ill-defined and loaded term.} is consistently long, say 3.7\,Gyr as occurred here. It has been estimated that Earth's biosphere will collapse as a result of the Sun's growing luminosity in approximately 0.9\,Gyr \cite{caldeira:1992}, which would mean that the latest epoch for life to start and still have time to lead to something like us would be 2.8\,Gya. In this picture, life \textit{must} start $(3.6\pm0.8)$\,Gya - else we would not be here to talk about it. Hence, the observed value of 3.7\,Gya is hardly surprising. This is an example of a selection effect influencing our bias, specifically a case of what has become known as the ``weak anthropic principle'' \cite{carter:1973}.

Bayesian statisticians have previously attempted to assess the case for a rapid abiogenesis, conditioned upon this chronology \cite{spiegel:2012}. However, that initial work did not allow both the abiogenesis and evolutionary timescales to be jointly inferred, which is critical given the covariance imparted by the weak anthropic principle. This author remedied this issue \cite{kipping:2020} and re-framed the problem in terms of hypothesis comparison (rather than posterior inference), which dissolves the influence of seemingly subjectively chosen priors on the emergence rates. The result is an expression for the \textit{minimum} odds ratio between a fast and slow abiogenesis scenario, given by

\begin{equation}
\frac{ \mathcal{Z}_{\mathrm{fast-life}} }{ \mathcal{Z}_{\mathrm{slow-life}} } \geq
\begin{cases}
\frac{ T }{ 2t_L' } & \text{if } T\geq2t_L'+t_I' ,\\
\frac{ T t_L' }{ 4(T-t_I')t_L' - 2t_L'^2 - (T-t_I')^2 } & \text{if } T<2t_L'+t_I',
\end{cases}
\label{eqn:1}
\end{equation}

where $T$ is the lifespan of Earth's biosphere, $t_L'$ is the time it took for the first evidence of life to appear in the historical record (which necessarily succeeds the true emergence date), and $t_I'$ is the time from then until intelligent life emerges (presumed to be the current epoch). Using previously published estimates of $T=5.304$\,Gyr \cite{wilde:2001,caldeira:1992}, $t_L'=0.7$\,Gyr from microfossils \cite{nutman:2016} and $t_I'+t_L'=4.404$\,Gyr (i.e. intelligence emerged recently) one obtains a 3.8:1 odds ratio. A 400\,Myr earlier start to life ($4.10\pm0.01$\,Gya) was indicated via the analysis of carbon-13/12 isotope ratios \cite{bell:2015}, and this adjusts the odds to 8.7:1. However, both fall short of the 10:1 threshold \cite{kass:1995} usually used to define ``strong evidence''. Accordingly, the evidence up to now has been tantalizing but not substantive.

\begin{figure}
\centering
\includegraphics[angle=0, width=16.0cm]{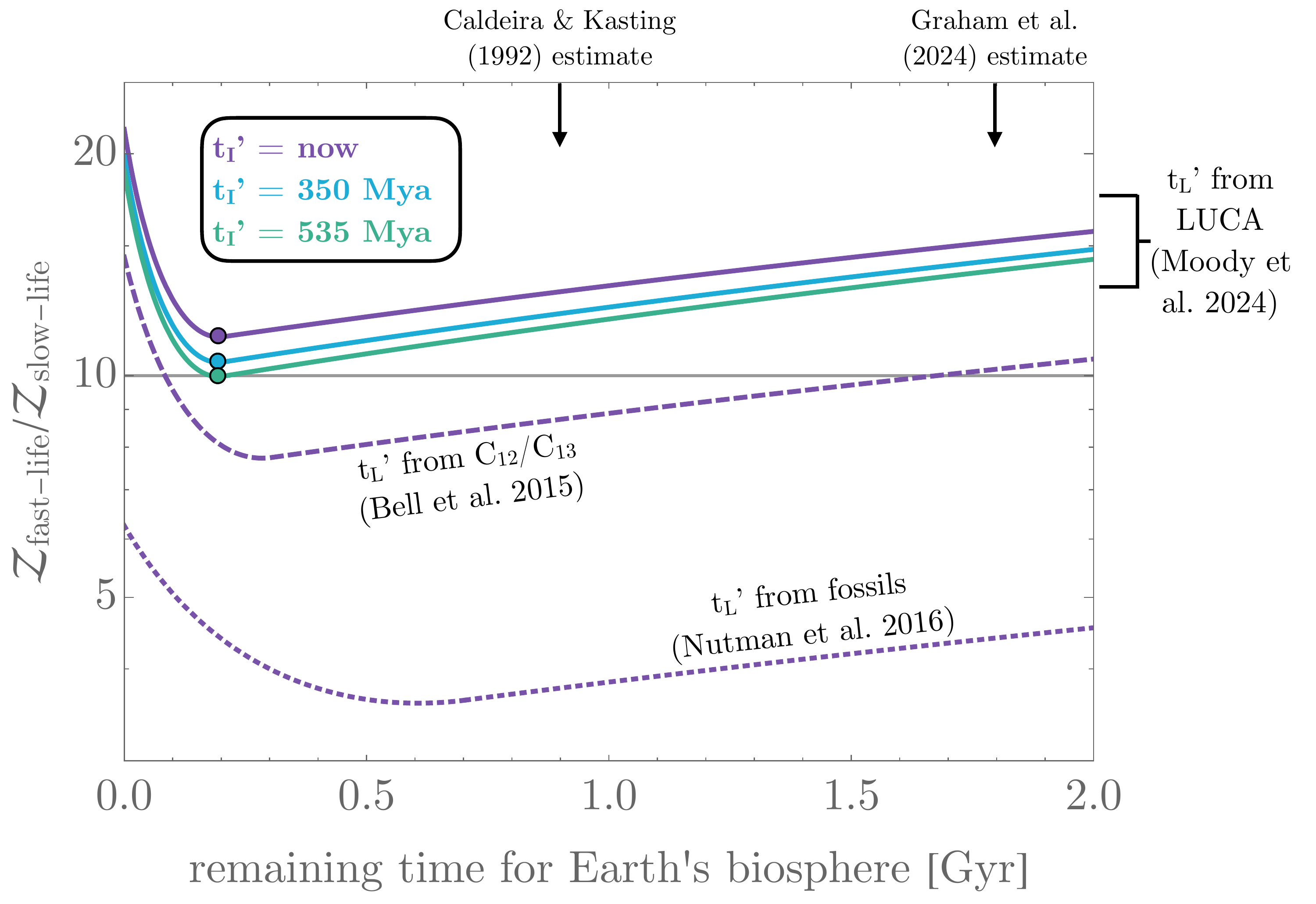}
\caption{\label{fig:1}
Odds ratio of the hypotheses that abiogenesis is a fast vs slow process.
Each curve represents the odds ratio, computed using Equation~\ref{eqn:1}, as a function of the biosphere's ultimate lifespan, $T$ - although we have subtracted off the modern epoch to yield remaining time on the $x$-axis. Previous estimates \cite{nutman:2016,bell:2015} for the earliest evidence of life yield the dotted and dashed lines, but the new LUCA date of 4.2\,Gya \cite{moody:2024} greatly improves the odds factor. Indeed, the three solid lines show how the odds are always above 10:1 irrespective of how one chooses $T$ or indeed fairly extreme choices (Carboniferious and Cambrian period) for the earliest civilization on Earth, $t_I'$ - evident by the circles which denote the minima.
}
\end{figure}


\section{The Implications of LUCA Living 4.2\,Gya}

A recent study \cite{moody:2024} revises these numbers by dating the Last Universal Common Ancestor (LUCA), using divergence time analysis of pre-LUCA gene duplicates and calibration using microbial fossils and isotope records. Updating Equation~(\ref{eqn:1}) to account for LUCA living $4.20_{-0.11}^{+0.13}$\,Gya \cite{moody:2024} (95\% interval) finally pushes us over the threshold to $13.0_{-3.1}^{+5.4}$:1 (68.3\% interval). However, the result is even more profound than this.

One might contest two numbers used in Equation~(\ref{eqn:1}): $T$ and $t_I'$. Concerning $T$, it was recently argued \cite{graham:2024} that the lifespan may be much greater than previously calculated (0.9\,Gyr), out to 1.8\,Gyr and this adjusts the odds to $15.2_{-3.6}^{+6.3}$:1. However, one might question whether a civilization could truly emerge during the tail-end of Earth's waning biosphere \cite{swansong:2013} and thus argue for a more conservative $T$. Figure~\ref{fig:1} shows the impact of varying $T$ in Equation~(\ref{eqn:1}). We certainly know $T\geq4.404$\,Gyr (since that is our arrival date) and thus one finds that in \textit{all cases}, the odds exceed 11.3 with a minimum at $T=4.6$\,Gyr (200\,Myr from now).

The result is even more robust than this, though. For one might also contest $t_I'$. The Silurian hypothesis suggests that a civilization could have emerged as early as the Carboniferous period (350\,Mya) and we would have no record of its existence today \cite{schmidt:2019}. Figure~\ref{fig:1} reveals that even this scenario leads to a minimum odds factor\footnote{Errors are not calculated here since the civilization age is now hypothetical.} of 10.4:1 for all possible $T$. Finally, perhaps the most extreme possibility is a civilization emerging during the Cambrian explosion (535\,Mya), the very start of complex life on Earth. Yet even this scenario still has a minimum odds factor of 10.0:1. The LUCA date is so far back that its effect overwhelms the weak anthropic principle, as well as any plausible variations on the other parameters involved.

Our work assumes that life began on Earth, rather than via panspermia. Consider instead that life began on Mars then transferred to Earth. The cooling timescale is proportional to planetary radius, and thus if Earth cooled from the Moon-forming impact (4.48\,Gya \cite{canup:2023}) to oceans in ${\sim}$76\,Myr \cite{wilde:2001}, then Mars could have cooled in 40\,Myr and had oceans no earlier than 4.50\,Gya. The latest (most pessimistic) abiogenesis date would be the LUCA value, giving $\mathcal{Z}_{\mathrm{fast-life}}/ \mathcal{Z}_{\mathrm{slow-life}} > 9.0$ assuming 0.9\,Gyr for Earth's habitability \cite{caldeira:1992}. This becomes 10.5 using the more modern 1.8\,Gyr value \cite{graham:2024}. Accordingly, despite invoking contrivances to engineer this number as small as possible, it still floats around the ``strong evidence'' threshold. For non-directed interstellar panspermia, it is shown in the appendix that this would require essentially all interstellar objects to be life-bearing in order to work, which is simply implausible.

There are two caveats to the main result presented. First, the LUCA date is a recent result that may not stand up to scrutiny; our result is conditional in this sense. Second, our result does not establish that life is common, since Earth's conditions could be incredibly rare \cite{ward:2000}. Our next task is clearly to look out and address this question: How common are conditions analogous to those of Earth?

\section*{Code and data availability}

No code was produced for this article.

\section*{Acknowledgments}
D.K. thanks donors
Douglas Daughaday,
Elena West,
Tristan Zajonc,
Alex de Vaal,
Mark Elliott,
Stephen Lee,
Zachary Danielson,
Chad Souter,
Marcus Gillette,
Tina Jeffcoat,
Jason Rockett,
Tom Donkin,
Andrew Schoen,
Reza Ramezankhani,
Steven Marks,
Nicholas Gebben,
Mike Hedlund,
Leigh Deacon,
Ryan Provost,
Nicholas De Haan,
Emerson Garland,
Queen Rd Fndn Inc,
Scott Thayer,
Ieuan Williams,
Xinyu Yao,
Axel Nimmerjahn,
Brian Cartmell,
Guillaume Le Saint,
Daniel Ohman,
Robin Raszka,
Bas van Gaalen,
Adam Taylor,
Josh Alley \&
Drew Aron. I am also grateful to organizers of the inaugural ``Astrobiology and the Future of Life Meeting'' in Houston, from which talks and conversations inspired this paper.

\providecommand{\noopsort}[1]{}\providecommand{\singleletter}[1]{#1}%

\clearpage

\section*{Appendix}

\label{si:minodds}

\section*{Minimum Odds Formula}

Let us here derive an expression for the minimum odds ratio, with respect to $T$, as shown by the three circles marked on Figure~\ref{fig:1}. It is first noted that that the phrase ``minimum odds'' has two contexts in this work. The first meaning is the original usage in the paper\cite{kipping:2020} that derived Equation~\ref{eqn:1}. In that work, ``minimum'' refers to the unknown parameter $\lambda_I$ - the rate of evolution to intelligent creatures. It was proven how, although we do not know what the true value of $\lambda_I$ is, we can always define a lower limit on $\mathcal{Z}_{\mathrm{fast-life}}/\mathcal{Z}_{\mathrm{slow-life}}$ by taking the limiting case of $\lambda_I \ll 1/t_I'$. This is what is meant by ``minimum odds'' in the original paper and explains why Equation~\ref{eqn:1} is an inequality, not an equivalence.

However, Equation~\ref{eqn:1} also has a minimum possible value with respect to the parameter $T$. This may be found by simply differentiating Equation~\ref{eqn:1} with respect to $T$ and re-arranging to make $T$ the subject, giving us a critical value of

\begin{align}
T_{\mathrm{crit}} &= \sqrt{t_I'^2 + 4 t_I' t_L' + 2 t_L'^2}.
\label{eqn:Tcrit}
\end{align}

Plugging the above into Equation~\ref{eqn:1} reveals the minimum (minimum) odds ratio, given by

\begin{align}
\frac{t_I' + 2t_L' + \sqrt{t_I'^2 + 4 t_I' t_L' + 2 t_L'^2}}{4t_L'},
\end{align}

which is used, along with Equation~\ref{eqn:Tcrit}, to define the locations of the minima (circles) in Figure~\ref{fig:1}.

\section*{ISO Estimates}

Consider a population of interstellar objects (ISOs) with a number density of $n$. The number of ISO impacts upon Earth per unit time will be this number density scaled by the cross-sectional area of Earth and the characteristic velocity of the ISOs, that is, $n v \pi R_{\oplus}^2$. We can refine this by increasing Earth's cross-sectional area to account for gravitational focusing - another factor of $(1+v_{\mathrm{esc}}^2/v^2)$. The characteristic timescale of such impacts is thus

\begin{align}
\tau \simeq \frac{1}{n v \pi R_{\oplus^2} (1+v_{\mathrm{esc}}^2/v^2)}
\end{align}

For ISOs, it is estimated that $n$ is ${\sim}0.1$\,AU$^{-3}$ \citep{laughlin:2017,trilling:2017,do:2018,moro:2018,levine:2021},
and the local stellar velocity distribution implies a characteristic $v \sim 30$\,km/s. Accordingly, one obtains $\tau \sim 200$\,Myr. Since LUCA appeared on Earth within 200\,Myr, panspermia via ISOs would require that essentially all ISOs are seeded with living organisms. 

\renewcommand{\refname}{Appendix References}

\providecommand{\noopsort}[1]{}\providecommand{\singleletter}[1]{#1}%

\end{document}